\documentclass[authorversion,nonacm]{acmart}

\AtBeginDocument{%
  \providecommand\BibTeX{{%
    \normalfont B\kern-0.5em{\scshape i\kern-0.25em b}\kern-0.8em\TeX}}}

\usepackage[english]{babel} 
\usepackage[utf8]{inputenc}
\usepackage{graphicx}
\usepackage{longtable}
\usepackage{tabularx}
\usepackage{multirow}
\usepackage{amsmath}
\usepackage{fancyvrb}
\usepackage{libertine}
\usepackage{zi4}
\usepackage{booktabs}
\usepackage{mathtools}

\DeclareMathOperator*{\argmax}{arg\,max}

\begin{document}

\title{On Prediction-Modelers and Decision-Makers: Why Fairness Requires More Than a Fair Prediction Model}

\author{Teresa Scantamburlo}
\authornote{Equal contribution.}
\email{teresa.scantamburlo@unive.it}
\orcid{0000-0002-3769-8874}
\affiliation{%
  \institution{Ca' Foscari University of Venice, European Centre for Living Technology}
  \country{Italy}
}

\author{Joachim Baumann}
\authornotemark[1]
\email{baumann@ifi.uzh.ch}
\orcid{0000-0003-2019-4829}
\affiliation{%
  \institution{University of Zurich, Zurich University of Applied Sciences}
  \country{Switzerland}
}

\author{Christoph Heitz}
\authornotemark[1]
\email{christoph.heitz@zhaw.ch}
\orcid{0000-0002-6683-4150}
\affiliation{%
  \institution{Zurich University of Applied Sciences}
  \country{Switzerland}
}

\begin{abstract}
An implicit ambiguity in the field of prediction-based decision-making regards the relation between the concepts of prediction and decision. Much of the literature in the field tends to blur the boundaries between the two concepts and often simply speaks of `fair prediction.' In this paper, we point out that a differentiation of these concepts is helpful when implementing algorithmic fairness. Even if fairness properties are related to the features of the used prediction model, what is more properly called `fair' or `unfair' is a decision system, not a prediction model. This is because fairness is about the consequences on human lives, created by a decision, not by a prediction. We clarify the distinction between the concepts of prediction and decision and show the different ways in which these two elements influence the final fairness properties of a prediction-based decision system. In addition to exploring this relationship conceptually and practically, we propose a framework that enables a better understanding and reasoning of the conceptual logic of creating fairness in prediction-based decision-making. In our framework, we specify different roles, namely the `prediction-modeler' and the `decision-maker,' and the information required from each of them for being able to implement fairness of the system. Our framework allows for deriving distinct responsibilities for both roles and discussing some insights related to ethical and legal requirements. Our contribution is twofold. First, we shift the focus from abstract algorithmic fairness to context-dependent decision-making, recognizing diverse actors with unique objectives and independent actions. Second, we provide a conceptual framework that can help structure prediction-based decision problems with respect to fairness issues, identify responsibilities, and implement fairness governance mechanisms in real-world scenarios.
\end{abstract}

\keywords{algorithmic fairness, prediction-based decision, responsibility, post-processing approach, social justice, human-in-the-loop, group fairness, decision systems, ethical decision-making, algorithm design}

\maketitle

\section{Introduction}

Algorithmic fairness has become a popular topic in the research community during the last years~\cite{barocas-hardt-narayanan,Kearns2019EthicalAlgorithm}, being increasingly addressed not only from a technical angle but also from a philosophical, political, and legal perspective~\cite{Binns2018,10.2307/24758720}. Algorithmic fairness is concerned with the consequences of prediction-based decisions on individuals and groups under the perspective of social justice~\cite{mulligan2019thing}. 
Since the beginning, the debate on algorithmic fairness has been focusing on the fairness of prediction models, which represent the core of Machine Learning (ML) research~\cite{pedreschi2008discrimination-aware,calders2010discrimination,Kamishima2012fairness-aware,Dwork2012,zemel2013learning-fair}. So it is not surprising that the focus of attention was put on how prediction models can create unfairness. 

We argue that the prediction model as such cannot be the reason for unfairness. It is the {\em usage} of the prediction model in its specific context which creates unfairness. For example, the recidivism risk model of the COMPAS tool~\cite{angwin2016machine} as such does not create racial discrimination. Only when it is used by judges who take decisions based on the COMPAS risk scores, such discrimination is created. Thus, the relationship between the properties of a prediction model, such as false-positive or false-negative rates, and possible harm for a specific group of the society, such as Afro-Americans in the case of COMPAS, rests upon an assumption of how the output of a prediction model creates actual consequences in the lives of people. 

This aspect is often neglected in the algorithmic fairness literature, assumptions on this relationship are often left implicit, and a fixed relationship between the prediction outcome and the impact on lives is taken for granted~\cite{Chouldechova2017,NIPS2017_a486cd07}. While assuming such a fixed relationship is convenient for studying the impact of the prediction model's features on the resulting fairness, it also ignores a central part of almost all implementations of influential prediction-based systems, which is the part of the actual decision-making: Only insofar as the output of a prediction model changes the course of the world, it can create unfairness. And {\em how} a prediction changes the course of the world depends strongly on how the prediction is actually used. 

As a prototypical case of how prediction models are implemented in real-world applications, we focus our discussion on {\em prediction-based decision systems}, where the outcome of ML prediction algorithms is used to make decisions affecting human subjects.\footnote{Such systems may be implemented either in the form of Automated Decision-Making (ADM) systems or in the form of a combination of a prediction system with a human decision-maker.}
We imagine a (human or automated) decision-maker who is taking decisions on people or for people, while this decision is informed by a prediction of some features of these persons. This is the typical scenario for many of the discussed cases of algorithmic fairness, such as a bank taking loan decisions based on repayment prediction, an enterprise taking hiring decisions based on job performance prediction, or a university taking admission decisions based on academic performance prediction\footnote{Note that other scenarios exist such as recommender systems where predictions are communicated to people who are taking decisions on themselves. In such cases, the findings of this paper are not directly applicable but they may inspire future work.}.

In such prediction-based decision systems, we may distinguish two conceptually different functions: First, we have the function of {\em prediction}, performed by a prediction model which is fed with individual data of a person, and whose output is some form of prediction of a target variable attributed to this person, which is not known to the decision-maker at the time of decision-making. This prediction might come in the form of a score, a probability, or a point prediction. Second, we have the function of {\em decision}, which is informed by the prediction, but in nearly all cases also influenced by additional parameters. For example, for a loan decision of a bank, not only the repayment probability but also the interest rate and the bank's business strategy may be decisive parameters. This idea has been studied in so-called cost-sensitive learning problems~\cite{elkan2001}. However, it remains unclear how the cost-sensitive approach changes once the additional requirement of fairness is introduced and how the concepts of prediction and decision interact in this process.

For studying the interaction of prediction and decision, we introduce a framework allowing us to distinguish the tasks and responsibilities of two different roles: The role of the `prediction-modeler,' and the role of the `decision-maker.' Following decision-theoretic concepts, we may think of two different agents, one being responsible for the prediction model and the other one being responsible for the decision-making. Our motivation for distinguishing these roles is not only fed by the theoretical analysis of how predictions are converted into (un)fair treatment as discussed above, but also by the observation that in practice these two roles are often split organizationally and covered by different people, different departments, or even different companies.%
\footnote{Of course, in an integrated and fully automated data-based decision system, both agents may be combined into one function, but we think that it is conceptually useful to distinguish the two functions.} Under a perspective of responsibility, the decision-maker is responsible for the decisions, and hence their consequences. However, as we will discuss below, the prediction-modelers also have their area of responsibility. They are responsible for creating the basis for a good decision, which consists in (a) delivering a meaningful and robust prediction (e.g., think of transparency and safety requirements in~\cite{hleg2019}), and (b) delivering all information needed for the decision-maker to care for fairness and other relevant ethical requirements (see accountability and fairness requirements in~\cite{hleg2019} and the obligations requested by~\cite{AiAct2021}). 

These two roles have different tasks and often conflicting goals. On the one hand, the prediction-modeler strives for prediction performance such as accuracy. This may be problematic when using ML models for consequential decision-making. For example,~\cite{Athey+2019} argues that standard ML prediction algorithms, optimized for accuracy, are not sufficient to take decisions in complex settings as there are often other relevant factors that are not represented in how well a model fits the training data. It also has shown that the fact that prediction-modelers usually have little specific knowledge of the domain in which an algorithm is applied may be problematic in consequential decision-making~\cite{athey2017beyond}. 
Similarly, \cite{cabitza2021accuracytrust} argue that optimizing for accuracy is imperfect and that a larger spectrum of metrics and information should be considered to assess a system's performance. On the other hand, the decision-maker aims to optimize their benefit resulting from the decision-making (e.g., considering business-related goals). These observations clearly show that the goal of a prediction-modeler needs not to be consistent with the final goal of the decision system.
The framework we propose addresses this tension. Specifying the prediction-modeler and the decision-maker as two different roles allows for separate performance measures.
Furthermore, it allows a separation of the responsibilities of the actors, which may also account for their domain-specific competencies. 

Our framework starts from a decision-theoretic analysis, thus connecting to existing literature that conceptualizes fairness as a decision-theoretic problem (see, for example,~\cite{petersen1976} and~\cite{sawyer1976}). However, instead of formulating fairness in terms of utility statements (e.g., see~\cite{10.1145/3097983.3098095} for a similar approach in the contemporary debate), we encode fairness as constraints of a decision problem.
This paper focuses on group fairness as the most established and most commonly used fairness category. This type of fairness intends to avoid systematic disadvantages of algorithmic decisions with respect to a sensitive attribute (such as gender, age, or race)~\cite{Binns2020,barocas-hardt-narayanan}.
There are also other types of fairness (for example, counterfactual fairness~\cite{NIPS2017_a486cd07}, individual fairness~\cite{Dwork2012}, or procedural fairness definitions~\cite{Grgic2019procedure}) but these are not covered in the present paper.

This paper is structured as follows: In Section~\ref{sec:Fairness-in-the-Machine-Learning-Literature}, we give a short review of the ML literature with respect to prediction-based decision-making, with a particular emphasis on group fairness metrics. In Section~\ref{sec:From-fair-predictions-to-fair-decisions}, we comment on the relationship between prediction and decision, which we then articulate and formalize in Section~\ref{s:Prediction-Based-Decision-Framework}. In Section~\ref{s:EthicalImplicationofOurFramework}, we discuss some insights derived from our framework. 

\section{Fairness in the Machine Learning Literature}
\label{sec:Fairness-in-the-Machine-Learning-Literature}

\subsection{Fairness of Prediction-Based Decision-Making Systems}
\label{ss:FairnessofAlgorithmicDecisionMakingSystems}

Prediction-based decision-making systems are increasingly used to assist (or replace) humans in making consequential decisions.
Algorithms are used to inform or automatically take decisions in lending~\cite{hardt2016equality,Fuster2017,Liu2018}, pretrial detention~\cite{angwin2016machine,Dieterich2016,Chouldechova2017,berk2021criminal,baumann2022sufficiency} college admission~\cite{Kleinberg2018}, hiring~\cite{Millera,Miller,Li2020,Raghavan2020}, insurance~\cite{baumann2023insurance}, and many other fields.
In recent years, we have seen a growing interest in the ethical implications of such prediction-based decisions, both from society and policymakers~\cite{CommissionAIAct2021}.
This has motivated the study of fairness in the field of ML, which has led to a newly formed community.%
\footnote{See for example: \url{https://facctconference.org/} and special tracks in major conferences in the field (ICML, NeurIPS, ECML, etc.).}

Many possible circumstances can lead to the development of algorithmic unfairness, such as a biased dataset, a systematic measurement error, the selection of a specific evaluation metric, or the taken modeling choices~\cite{mitchell2021algorithmic}.
In fact, biases can arise throughout many different stages of the ML-based decision-making life cycle~\cite{Suresh2021SourcesOfHarm,BiasOnDemand2023FAccT}.
Pursuing the goal of alleviating issues of algorithmic unfairness, researchers have proposed a plethora of fairness definitions~\cite{narayanan2018translation,verma2018fairness}.
In this paper, we focus on group fairness, a definition that has been of particular interest in the literature on fair ML~\cite{pessach2020algorithmic}.
We now introduce the most common group fairness metrics before we describe how they can be ensured.

\subsection{Measuring Group Fairness}
\label{ssec:Measuring-Group-Fairness}

We use $A$ to denote the \textit{sensitive attribute} (sometimes also referred to as \textit{protected attribute}).
Following related work, we consider binary group membership $A=\{0,1\}$, but our arguments generalize multi-group situations.
$\vec{X}$ denotes the observable attributes that are used for prediction%
\footnote{
Notice that $\vec{X}$ may or may not contain $A$.
Not using the sensitive attribute as an input for the ML algorithm refers to a somewhat naive concept of fairness called \textit{fairness through unawareness}~\cite{grgic2016case}, which does not effectively avoid disparate impact in case of redundant encodings (meaning that the sensitive attribute can be predicted by the remaining observable attributes, which is a likely scenario in the age of big data)~\cite{pedreschi2008discrimination-aware}.
}, while $Y$ denotes the unknown but decision-relevant target variable. For the sake of simplicity, we assume $Y$ to be a binary variable. We assume that there is \textit{prediction function} $f$ that maps instances of $\vec{X}$ to a prediction $\hat{Y} = f(\vec{X})$.
The \textit{decision function} is a (possibly group-specific) function $d(\hat{Y})$ or $d(\hat{Y}, A)$ that transforms the prediction $\hat{Y}$ into a decision $D$.

According to~\cite{barocas-hardt-narayanan} and~\cite{Kearns2019EthicalAlgorithm}, most of the existing group fairness criteria fall into one of three categories: independence, separation, or sufficiency.
Due to ambiguities regarding the notion of a `fair prediction' vs. that of a `fair decision,' different notations are being used for the same fairness criterion.
Those who apply the criteria to prediction models usually refer to the prediction $\hat Y$ (sometimes also expressed as a score, usually denoted by $S$ or $R$), while those applying it to decision algorithms refer to the decision $D$ for the same criteria~\cite{hardt2016equality,verma2018fairness}.%
\footnote{
An example of such a discrepancy in notations is the group fairness metric called separation (which we introduce shortly) as defined in the algorithmic fairness literature.
For example,~\cite{hardt2016equality} defines this notion of fairness for a sensitive attribute $A = \{0,1\}$ as: $P(\hat Y=1|A=0,Y=y)=P(\hat Y=1|A=1,Y=y),y \in \{0,1\}$.
This definition implicitly assumes that a specific value of the prediction $\hat Y$ is equivalent to a specific decision.
Others, such as~\cite{verma2018fairness}, define separation for a sensitive attribute $G = \{m,f\}$ and a decision $d$ as follows: $P(d=1|Y=i,G=m)=P(d=1|Y=i,G=f),i \in 0,1$.
}
In this paper, we use the latter notation (which is also used by~\cite{verma2018fairness} and~\cite{mitchell2021algorithmic}, for example) because it is in line with the framework we propose.

\textit{Independence} (also called \textit{statistical parity}, \textit{demographic parity}, or \textit{group fairness}) requires the decision to be independent of the sensitive attribute and is formally defined as:
\begin{equation}
    P(D=d|A=1)=P(D=d|A=0).
\end{equation}
Thus, the probability of a specific decision $d$ must not depend on the group membership $A$.
For the example of granting a loan, independence requires equal acceptance rates for both groups. 
\textit{Conditional statistical parity} extends independence in that it allows a set of legitimate features $L$ to affect the decision~\cite{Kamiran2013,10.1145/3097983.3098095}:
\begin{equation}
    P(D=d|L=l,A=1)=P(D=d|L=l,A=0).
\end{equation}
For example, in the loan case, the applicant's requested credit amount could be a possible legitimate feature.

\textit{Separation} (also called \textit{equalized odds}) takes the individual's $Y$-value into account:
\begin{equation}
    P(D=d|Y=y,A=1)=P(D=d|Y=y,A=0).
\end{equation}
Thus, the requirement of the same probability of a decision $d$ across groups is restricted to individuals with the same value of $Y$. Separation is equivalent to parity of true positive rates (TPR) and false positive rates (FPR) across groups $a \in A$.
Another popular definition of fairness, \textit{equality of opportunity}, is a relaxation of the separation constraint only requiring TPR parity.%
\footnote{
Similarly, FPR parity (also called \textit{predictive equality} by~\cite{10.1145/3097983.3098095}) is a relaxation of separation that only conditions on $Y=0$.
}
In the loan granting scenario, this definition of fairness would ensure that ``deserving individuals'' (the ones who would repay the loan if given one, i.e., $Y=1$) receive loans proportionately across groups.

In contrast, the fairness notion \textit{sufficiency} conditions not on $Y$ but on the decision $D$:
\begin{equation}
    P(Y=y|D=d,A=1)=P(Y=y|D=d,A=0).
\end{equation}
This means that, among all those individuals who receive the same decision $d$, the probability of a specific value $y$ must not depend on $A$. For binary $Y$ and $D$, sufficiency is equivalent to parity of positive predictive values (PPV) and false omission rates (FOR) across groups -- meaning that for subgroups formed by $D$, an equal share of individuals must belong to the positive class $Y=1$ across groups $A$~\cite{baumann2022sufficiency}.
The fairness definition PPV parity (also called \textit{predictive parity} by~\cite{Chouldechova2017,Kasy2021}) relaxes sufficiency in that it only requires $Y$ and $A$ to be independent for all individuals who received a positive decision $D=1$, which amounts to parity of PPV for binary classification~\cite{baumann2022sufficiency}.%
\footnote{
Similarly, FOR parity is a relaxation of sufficiency that only considers individuals who received a negative decision $D=0$~\cite{baumann2022sufficiency}.
}

Concluding, we see that all group fairness definitions are based on the equality of a specifically defined probability across the considered groups.
Interestingly, the ML literature does not systematically relate the equality of these probabilities to philosophical concepts of social justice and fairness.
However, it is beyond the scope of this paper to dig into this question of the relation of the mathematical definition of fairness metrics and their moral meaning, even though this is still only rarely discussed, for example, in~\cite{hedden2021statistical,Loi2019,baumann2022SDS_fairness_principle,long2021fairness,Hertweck2021}.
For the current paper, it suffices to state that measures of group fairness are typically based on the equality of conditional probabilities, which corresponds to the normative idea of `equal shares' across the different groups.

\subsection{Generating Group Fairness}
\label{ssec:Implementing-Group-Fairness}

The context-dependent nature of the fairness problems makes it impossible to agree on one universally applicable definition of group fairness. In addition, many fairness criteria are mathematically incompatible~\cite{Kleinberg2016,Chouldechova2017,garg2020fairness,Friedler2021}. This requires making a choice based on the concrete setting of the decision problem. 

There are different techniques for ensuring the fairness of prediction-based decision systems, most of which fall into one of three categories~\cite{mehrabi2019survey}:
\textit{Pre-processing} describes a method in which the training data is manipulated in order to generate a prediction model that leads to a fair decision~\cite{calders2010discrimination,kamiran2012dataprocessing}.
\textit{In-processing} refers to implementing fairness requirements into the prediction model itself (e.g., by incorporating a fairness constraint for the training of a prediction model)~\cite{Kamishima2012fairness-aware,BilalZafar2017}. 
The third category, \textit{post-processing}, takes the prediction model as given and changes the decision function such that the resulting decisions meet some fairness constraints (e.g., by using group-specific thresholds on predicted scores)~\cite{baumann2022sufficiency,hardt2016equality,pmlr-v81-menon18a,10.1145/3097983.3098095}.
Despite their inherent advantages and disadvantages, all of these methods have been shown to be effective~\cite{barocas-hardt-narayanan}.%
\footnote{We point to~\cite{pessach2020algorithmic,caton2020fairness} for a more detailed discussion of the different unfairness-mitigation techniques including their (dis)advantages.}
Pre-processing and in-processing techniques place the burden of generating fairness on the prediction-modeler. Conceptually, this is only possible if the decision-maker is not an independent actor but instead implements a predefined decision rule applied to the output of the prediction model. In Section~\ref{sec:From-fair-predictions-to-fair-decisions}, we will point out that this is an unrealistic assumption in many cases. The much more frequent situation is that a prediction model may be used in different ways by an independent decision-maker, who, e.g., considers additional factors for the decision-making. This implies a focus on post-processing methods.

\section{The Relation between Predictions and Decisions}
\label{sec:From-fair-predictions-to-fair-decisions}

In popular narratives of algorithmic decision-making, the distinction between the idea of decision and that of prediction seems to be blurred and applied to the notion of fairness in a flexible way. Neologisms like ``fair prediction''~\cite{Chouldechova2017} or ``fairness-aware learning''~\cite{Kamishima2012fairness-aware} have become familiar within the ML community fueling, often unintentionally, the idea that fairness is a property of a prediction model. Even studies addressing algorithm-human interaction ultimately assimilate human decisions to a prediction task, e.g by comparing human estimates to algorithmic outcomes~\cite{kleinberg2018human, dressel2018accuracy, vaccaro2019effects, green2019disparate}.

Actually, the collapse between the two concepts does not reflect an explicit ideological position and some studies clearly specify that fairness is an attribute that refers to a decision rule~\cite{corbett2018}. However, formal characterizations tend to apply fairness criteria to the prediction model (e.g., the classifier), assuming that the decision consists of the prediction outcome (e.g., see~\cite{zafar2015, pmlr-v81-menon18a, berk2021criminal}). 

Given similar formulations, one might naturally assume that the relation between prediction and decision is fixed and given, i.e., that a specific prediction leads to a specific decision. However, this is not true in many realistic examples, where the optimal decision depends on the prediction and other parameters (for example, in lending decisions, on the interest rate). This is in line with the idea of cost-sensitive learning~\cite{elkan2001}. Thus, qualifying a prediction as fair is misleading unless we explicitly assume how a prediction is converted into a decision. In general, the fairness attribution applies more properly to the full system (i.e., the combination of prediction and decision rule) rather than to the prediction as such. 

In a similar vein, \cite{kupplerfair} distinguish between prediction and decision in data-driven decision procedures to highlight the different meanings and roles of fairness and justice. According to the authors, (algorithmic) fairness is concerned with the statistical properties of the prediction model, whereas justice is concerned with the allocation of goods and, therefore, more appropriately associated with the decision step. Notice that our approach is different as we build on the idea that fairness is a concept related to the outcomes of decisions in people's lives. Therefore, we argue that fairness is a property of the entire system and that theories of distributive justice should be reflected in the chosen fairness definition. Instead of fully disentangling the desired properties of a prediction model from the decisions step, we argue that the prediction model's sole purpose is to inform decision-makers. This allows for building prediction-based decision-making systems around social fairness desiderata, including theories of distributive justice that are morally appropriate for the context at hand.  

\subsection{Why a Distinction is Needed}
Abstract formalization facilitates the overlap between the concepts of prediction and decision. For example, in classification tasks, the goal of prediction is to select an option among possible alternatives so that predicting can be viewed as a special form of deciding. Also, in cases where the outcome to be predicted is a numerical value (e.g., a risk score), a prediction can be easily translated into a discrete scale (e.g., low - medium - high risk). From this standpoint, there is not much difference between the task performed by a prediction model and that performed by a decision-maker. However, if we go beyond mathematical abstractions and take an ethical stance, a decision is not just a matter of choosing among alternatives. It is a way to act and impinge upon humans and the environment. In other words, decisions change the \emph{status quo}, thus bearing consequences for the decision-maker, the decision subjects, and possibly the external world. On the contrary, a prediction, per se, has no impact, and its capacity to influence decision-making is made possible only by a policy or a decision rule. It is the latter that specifies the consequences of future actions. 

Consider, for example, a bank giving loans to individuals, building their decision of accepting a loan applicant on a predicted repayment likelihood. Granting a loan creates a tangible impact in the form of a benefit consisting of improved financial flexibility and new buying options. This benefit is denied to loan applicants who receive a negative decision. Apparently, the prediction algorithm influences the decision, but the prediction itself is not what creates (un)fairness, it is the decision specifying how to use the prediction estimate. Note that even if the decision is fully determined by the prediction -- a case which is rarely met -- the question of whether the prediction algorithm is fair or not is conditioned on the assumed relation between prediction and decision rule. This is why we conceptually suggest to clearly distinguish between the two elements of prediction and decision, which both are ingredients of any prediction-based decision system, whether it be fully automatic or also influenced by humans.
Most importantly, the distinction between the two concepts invites us to contextualize algorithmic decision-making into a process of social construction reflecting value judgments and power asymmetries.

Often, apart from the prediction itself, the final outcome of a decision process is determined by additional pieces of information. Consider, again, a bank that needs to decide whom to grant and whom to deny a loan, where the bank's goal is to maximize their profit from the loan business. It is clear that the expected profit depends on the probability of repaying, which is why a prediction model for determining this probability is needed. However, other parameters, such as the interest rate charged for the loan, are also relevant, and these parameters obviously determine the minimum repaying probability that the bank will accept. A change of this threshold changes the decision rule and, thus, this represents a cost-sensitive learning problem~\cite{elkan2001}. However, what is not covered in the literature on cost-sensitive learning is the fact that this often also changes the decision system's fairness properties. If the decision is fair for one threshold, this does not imply that it is fair for another threshold.

Another reason for marking a distinction between prediction and decision lies in the fact that the two concepts are benchmarked against different criteria. From a conceptual point of view, independent of the decision that may follow, a prediction can only be assessed in terms of its predictive power, e.g., accuracy. If one predicts, for example, the probability that a loan applicant will repay their debt, then a given prediction algorithm can be more or less accurate, typically evaluated with observation data, which are called ``ground truth.'' It is, conversely, nonsensical to ask whether a decision is accurate or not since, broadly speaking, there is no such thing as ``ground truth'' in a decision process. A decision can be ``right'' or ``wrong,'' but the same decision can be qualified differently depending on a variety of factors. We can judge the quality of a decision based on the purpose it aims to achieve and the consequences it has on the decision-maker and their surrounding environment (including other people), for example, in terms of fairness and accountability. As we will see, decision theory frames this intuition as an optimization problem so that an optimal decision is the one that maximizes a specific goal.

The problem of whether a prediction can be seen as unfair or not connects to a broader philosophical debate. In particular, this issue recalls the attempt to investigate the moral status of beliefs and thoughts going beyond the sphere of actions and deliberations.
For example, advocates of epistemic injustice argue that people can commit injustice when they fail to believe someone's testimony due to prejudice~\cite{fricker2007epistemic}, and theorists of doxastic wronging postulate that people can wrong others in virtue of what they believe about them, and not just in virtue of what they do~\cite{basu2019wrongs}. Therefore, one may ask whether a (un)fair prediction could be regarded as an unjust or discriminatory belief.
In this paper, we do not dig into this problem, which would require a separate discussion, and consider unjust beliefs on par with a decision rule operating, more or less consciously, in the decision-maker's mind.

\subsection{Related Studies on the Interaction with Prediction-Based Decisions}

Arguing that the assessment of fairness requires a distinction between prediction and decision recalls a growing body of research focusing on how humans and algorithms interact when making decisions. These studies approached the interaction from different perspectives. 

Some works aimed at showing how algorithms can improve predictive performance~\cite{kleinberg2018human, miller2018want} especially when there are carefully designed protocols of interaction~\cite{cabitza2021studying}. Others investigated people's perceptions, shedding light on what has been called ``algorithmic aversion''~\cite{dietvorst2015algorithm}, i.e., the situation in which a human decision-maker prefers human forecast over algorithmic prediction even if the latter is more accurate than the former. Further research pointed out that human decision-makers tend to deviate from algorithmic predictions~\cite{stevenson2021algorithmic} and struggle to assess algorithmic performance~\cite{green2019principles, poursabzi2021manipulating}. Similar works suggested best design practices to allow designers to make adjustments in fully automated decision systems that interact with people (the so-called ``street-level-algorithms'') and make erroneous or unfair decisions when encountering a novel or marginal case~\cite{alkhatib2019street}. 

Another area of research focuses on the challenge of autonomy in algorithmic decision-making. A key aspect of this work involves clarifying the meaning of autonomy when referring to algorithms, making distinctions between being ``autonomous'' and being ``automatic'' \cite{chiodo2022human, pianca2022interdependence}. Further concerns regard the constraints that algorithms may pose to decision-makers' agency (e.g., in relation to the algorithm's perceived authority) \cite {hayes2020algorithms} or the liability of algorithms causing injuries to humans or property damages \cite{barfield2018liability}.

Our work complements this broader literature and suggests new research directions exploring the interaction between the actors who deal with the prediction and the decision tasks. In this way, we aim at gaining a better understanding of the context of prediction-based decision systems, suggesting the fundamental social nature of systems' construction process and highlighting the informational gaps that must be addressed to improve the accountability of prediction-based decision-making systems.

\section{A Prediction-Based Decision System Under Fairness Constraints}
\label{s:Prediction-Based-Decision-Framework}

Intuitively, a decision is a termination of a process that involves several tasks. In general, a decision-maker identifies their preferences, sets up requirements and courses of action, analyses the pros and cons of each alternative, and selects one among possible options -- the etymology of the term is quite explanatory: from Latin ``de'' = ``off'' + ``caedere'' = ``cut.'' 

The idea that distinct tasks are involved in human decision-making is well-entrenched in classical philosophy. Medieval philosophers, for example, acknowledge three distinct operations (see e.g.~\cite{Aquinas} and~\cite{saarinen2006weakness, hain2015consilium} for a modern interpretation). The first one, the so-called ``consilium,'' consists of asking for advice and gathering relevant information for the decision at stake. The second operation involves judgment and constitutes more properly the deliberation step, also known as ``resolution.'' The third operation regards the concrete actions that implement what was decided in the previous step.\footnote{The conceptualization of prediction-based decision as a two-step process has another parallel with the philosophy of science, where a famous distinction regards the generation of new knowledge (the ``context of discovery'') and its assessment (the ``context of justification''). In this work, we recalled the analogy with classical moral philosophy but we acknowledge that the parallelism with the pair ``discovery-justification'' would offer other important stimuli that would deserve e dedicated discussion.}

Our framework focuses on the first two operations of this deliberation scheme (``seeking advice'' and ``deciding''), but since we consider decisions that take place in uncertain conditions, we will focus on a specific type of advice generated by a prediction model. In addition, our framework assumes that a specific ethical constraint about group fairness (already described in Section~\ref{ssec:Measuring-Group-Fairness}) is imposed. 

First, we introduce two formal roles: the role of the prediction-modeler and the role of the decision-maker. We show how these connect to different goals and tasks of prediction-based decision systems and where fairness constraints (FC) come into play. Then, we describe the two roles in a more formal way, specifying the parameters that characterize the tasks of both actors. Finally, we specify how these two interact and, in particular, define a minimum set of deliverables required to construct optimal decision rules meeting fairness constraints.

\subsection{Two Roles in Prediction-Based Decisions}

In our framework, we distinguish two roles that become particularly relevant in the discussion of responsibilities connected to a prediction-based decision system. The first role is the prediction-modeler, who cares for the prediction. The second one is the decision-maker who uses the prediction to optimize their own benefit (utility) while possibly also considering fairness issues by ensuring that certain fairness measures are met. Note that following the analogy with the aforementioned three-step model of decision-making, the role of prediction-modeler and that of decision-maker fulfill, respectively, the tasks of advising and deliberating.\footnote{Here, we present a simplified characterization focused on two roles but more sophisticated descriptions could rely on multiagent system theory \cite{singh1994multiagent}}.

Usually, these two roles reflect different backgrounds and often different education. Typically, the role of the prediction-modeler is taken by data scientists, engineers, or computer scientists, while the role of the decision-maker can be played by various professionals, such as doctors, product managers, or business strategists. In the bank setting, the prediction-modeler may coincide with an external and independent organization (say, a software company) or an internal but separate department (e.g., the bank's data science lab), while bank managers play the role of the decision-maker.
The source of the distinction between the two roles lies in the different goals they aim to achieve. While the goal of the prediction-modeler is to maximize the performance of a prediction model, such as accuracy, the goal of the decision-maker may vary depending on the context and includes, for instance, the increase of profit or the optimization of product development. 

Our framework rests on the idea that even though the two roles are conceptually and practically distinct, they need to work in synergy for addressing fairness issues. Our framework specifies the tasks related to each role and, at the same time, the interaction points that allow the decision-maker to adequately integrate fairness concerns into decision-making (see Sections \ref{ssec:The-decision-maker}-\ref{ssec:The-interaction}).

The decision-maker is the role directly involved in choosing which fairness metric to use (i.e., how unfairness is measured) and to what extent unfairness should be removed.
These choices require the assessment of the social and the business context of the decision system. Typical questions to be answered are: Which subgroups should be considered with respect to fairness (i.e., what are the sensitive attributes)? Which fairness metric is the most appropriate in the given social context? What is the optimum trade-off between optimizing utility and enforcing fairness? 

Answering these questions is hard, if not impossible, for the prediction-modeler whose task is predicting an unknown but decision-relevant quantity $Y$.%
\footnote{Note that the notion of a prediction in the context of an ML model is more encompassing than it is in colloquial language. While, in everyday language, we use the term ``prediction'' to refer to future situations (e.g., whether it will rain tomorrow), in the field of ML and statistics, a prediction simply relates to a fact that is not known when taking some action (e.g., ``whether patient x has disease y`` or ``whether applicant z is trustworthy or not''). This lack of knowledge may be caused by different reasons, for example, due to missing information, but also when referring to an event in the future.} On the one hand, one may argue that, in principle, the prediction task should not involve caring for fairness-relevant issues: A good prediction is something else than fair treatment or a socially just distribution of benefits and harms. So, from a conceptual point of view, one may question whether assigning responsibility to the prediction-modeler makes any sense. On the other hand, from a practical perspective, the prediction-modeler is often simply not able to care for fairness because they do not have access to the needed contextual information and do not have the competence to decide on the normative issues involved. This makes clear why, both from a conceptual and a practical viewpoint, the two roles should be distinguished and why these are separated in most real-world cases. 

Note that these roles are often left implicit in most ML fairness literature, where the common narrative of ``fair ML'' or ``fair prediction models'' would indirectly suggest that caring for fairness is a task of ML engineers. Our framework aims to be more specific than the standard approach in defining the roles and the minimum requirements associated with these roles in prediction-based decision-making. This will allow us to derive ethical responsibilities and support the implementation of fairness governance mechanisms in real-world scenarios. 

In the following subsections, we will analyze the two roles and their interaction more closely, which will lay the ground for answering the question of who is responsible for what. 

\subsection{The Decision-Maker}
\label{ssec:The-decision-maker}

Decision-making is a task that can be described in purely abstract terms. This is what decision theory does to frame a variety of decision problems ranging from what movie to watch in the evening to what career to pursue after college.

We consider a decision-theoretic agent%
\footnote{We are aware that the decision-maker (also called the rational agent) assumed in economic theory is an idealization that might be far away from reality (i.e., humans can make irrational decisions for many reasons), but a decision-theoretic approach can also be a good starting point for modeling and analyzing decisions and their consequences.}
who makes decisions based on certain goals and preferences. In its simplest form, the agent chooses an action in a finite set of possible alternatives, and this action has an impact on the surrounding environment. For evaluating the impact, we consider the system's state after the agent's action and assign each possible state a specific value of the so-called \emph{utility}. This refers to a quality that measures the desirability of this future state: the more desirable the state, the higher the utility. Thus, utility formalizes and quantifies the notion of a goal. It allows comparisons among different future states as a function of the chosen action, which, in turn, allows one to choose among the different possible actions. In many cases, the relation between action and outcome (and thus utility) is not deterministic: The same action might lead to different outcomes, depending on factors that are not under the decision-maker's control. This situation is referred to as a ``decision under uncertainty.'' It puts a decision-maker in a situation where they have to make a decision without really knowing what utility will be realized. In other words, the utility achieved following a decision is a random variable. Decision-making under uncertainty is about managing this uncertainty while still trying to achieve a goal. 

In the loan example, there are two possible future states or outcomes at the end of the loan contract: The loan plus associated interests may be paid back, or the debtor has defaulted, resulting in a loss of the loan. Obviously, for the bank, the former state is more desirable than the latter. The utility can be measured, e.g., by the amount of money that the bank has in their accounts by the end of the contract duration. 

For applying this general decision-theoretic framework to the case of prediction-based decision systems, we identify the concept of ``action'' with that of ``decision.'' The uncertainty of the outcome is usually attributed to the lack of knowledge of a random variable $Y$, which might take different values $y$. Note that in real-world situations, many other factors might create uncertainty, but in the following, we analyze the simplest case in which $Y$ is the only source of uncertainty. In the loan example, $Y$ corresponds to the repayment of the debt by the debtor, which decides which state is reached at the end of the loan contract. We also assume that the decision-maker takes not only one single decision but a sequence of many decisions of the same kind, which is a standard assumption for prediction-based decision systems. In the loan example, we envision a sequence of loan decisions of the bank, following the same decision rules for acceptance. 

In such a situation, the goal achievement is measured as an expectation value, i.e., the decision-maker is interested in a decision rule that creates maximum utility in the long run, which means that they try to maximize the \emph{expected utility} $E(U)$ as a function of their decision $D$: 
\begin{equation}
    E(U(D))=\sum_s P(s|D)\cdot U(s|D) 
\end{equation}
where each state $s$ represents a possible outcome as a function of the decision $D$, and $U(s|D)$ are the utilities associated with each outcome $s$. If $Y$ is the only source of uncertainty, and thus determines the outcome, the different outcome states $s$ correspond to the different values $y$ of the random variable $Y$: 
\begin{equation}
\label{expected_utility}
    E(U(D))=\sum_y P(Y=y|D) \cdot U(Y=y|D) 
\end{equation}
where now $U(Y=y|D)$ denotes the utility for the state reached in case of $Y=y$, and $P(Y=y|D)$ is the probability that this state is reached. Note that both elements may depend on the decision $D$.  

For the sake of simplicity, in the following, we restrict ourselves to a binary variable $Y$, with two values $y=0$ and $y=1$. This gives:
\begin{equation}
    E(U(D))= P(Y=1|D) \cdot U(Y=1|D) + (1-P(Y=1|D))\cdot U(Y=0|D) 
\end{equation}
A decision-maker would be called rational if they choose the action that maximizes their expected utility (see the principle of Maximum Expected Utility~\cite{russell2010artificial}): 
\begin{equation}
\label{optimize_utility}
   D = \argmax E(U(D))
\end{equation}

For a simple loan example, the decision $D$ is binary, with $D=1$ corresponding to accepting the loan.\footnote{
Notice that decision rules are likely to be more complex in reality. For example, a bank could adjust the interest rate depending on the predicted repayment probability and the applicant's willingness to pay (only denying a loan in cases with a very high default probability).
Our framework generalizes to more complex decision rules. However, for simplicity, we consider a binary case.
}
If we set the repaying probability $p=P(Y=1)$, then this reads:
\begin{align*}
    E(U(D=1)) & = p \cdot \alpha - (1-p)\cdot \beta \\
    E(U(D=0)) & = \gamma
\end{align*}
where $\alpha$ is the profit of the bank if the customer pays back, $\beta$ is the loss if the customer defaults, and $\gamma$ is the profit that can be made by not giving the loan but instead investing the money into another business line of the bank.  

The optimization problem with respect to the decision (see Eq. (\ref{optimize_utility})) can easily be solved, leading to:
\begin{equation}
\label{optimum_solution_loan}
    D=1 \text{\;\;if\;\;}  p>\frac{\beta+\gamma}{\alpha+\beta}, \text{\;\;\;\;} D=0 \text{\; else.} 
\end{equation}
This example shows that the decision rule depends not only on the probability $p$, but also on other parameters ($\alpha,\beta,\gamma$) which are independent on the prediction of $Y$ (given by $p$), but still decision-relevant.%
\footnote{
We assume $\alpha + \beta \neq 0$, as otherwise there would be no need for a prediction in the first place. In the loan example, if $\alpha = -\beta$, the bank's profit would be the same for any value of $D$.
}
In line with cost-sensitive learning approaches~\cite{elkan2001}, this exemplifies why the prediction alone does not solve the decision problem. 

An important insight from this decision-theoretic analysis is that the decision-maker needs the probabilities $P(Y|D)$ to optimize their decisions, which directly leads to the necessity of a prediction model. In fact, the fundamental equation (\ref{expected_utility}) is composed of two elements: the probabilities $P(Y|D)$ and the utilities $U(Y|D)$. The first element is the one that is related to the prediction task, and the second element is related to the decision context, as it implements the desirability of the different possible outcomes. Both elements are independent of each other. 

Up to now, we assumed that the decision-maker bases their decision strictly on maximizing their utility. The resulting optimum decision rule, given by the solution of Equation (\ref{optimize_utility}), may or may not produce fairness issues. A decision-maker who also wants to consider fairness in their decision-making has to adopt their decision strategy such that the resulting decision fulfills the chosen fairness criterion.
While many different ways of how to do this are conceivable, a natural way of extending Equation (\ref{optimize_utility}) to a fairness-sensitive context is to impose a fairness constraint on the utility maximization: 

\begin{equation}
 \label{optimize_utility_fair}
 \begin{split}
   D = & \argmax E(U(D)) \\
   \text{subject to }  & \textit{Fairness Condition FC}
\end{split}
\end{equation}
where $FC$ is a condition of equality such as the ones mentioned in Section~\ref{ssec:Measuring-Group-Fairness}, or a relaxed version of them. From a formal decision-theoretic perspective, this is the optimal combination of the decision-maker's original goal and the additional consideration of fairness.

In the context of a post-processing approach for ensuring fairness, this constraint optimization problem has been solved in~\cite{hardt2016equality} (for the fairness metrics equalized odds, equality of opportunity, and predictive equality), in~\cite{10.1145/3097983.3098095} (for the fairness metrics statistical parity and conditional statistical parity), and in~\cite{baumann2022sufficiency} (for the fairness metrics sufficiency, predictive parity, and FOR parity).

\subsection{The Prediction-Modeler}
\label{ssec:The-prediction-modeler}

As illustrated in the preceding subsection, the decision-making process requires the probabilities $P(Y|D)$ to solve problems that involve an unknown quantity $Y$. In the context of machine learning, this corresponds to making a probabilistic prediction of $Y$ that the decision $D$ might depend on. This represents the prediction task that the prediction-modeler is expected to address.

Interestingly, this does not include all versions of prediction models used in ML and discussed in the context of fairness. In particular, a point estimator $\hat Y$ with two possible values $\hat Y=\{0,1\}$ is of little use for the decision-maker, as this does not allow to solve the decision problem stated in Eq. (\ref{optimize_utility}). Consider, for example, the optimum solution (see Eq. (\ref{optimum_solution_loan})) for $p \gg 0.5$ and for realistic parameters $\alpha$ and $\beta$: a typical ML prediction model optimized for maximum accuracy (threshold $p=0.5$) would lead to far too many instances of $\hat Y=1$ and thus $D=1$. In general, we can conclude that a prediction-modeler who does not have access to the external parameters $\alpha,\beta,\gamma$ is not able to deliver a good point estimator $\hat Y$, which allows solving the decision-maker's task.

A typical assumption in the ML (fairness) literature is that the decision is determined by the value of $Y$ (see~\cite{murphy2012machine,hardt2016equality,mitchell2021algorithmic}), e.g., such that $Y=1$ implies $D=1$, and vice versa. This means that if only $Y$ can be predicted with high accuracy, then the decision $D$ will be correct. Sometimes it might be possible to achieve a perfect prediction, e.g., in the case of picture recognition. Here, the fact that a picture represents a dog instead of a cat is evidence that could be checked at the time of decision-making (or recognition), even if an ML classifier does not predict the image correctly. However, this is not the case in many decision problems discussed in the algorithmic fairness literature. For instance, for the loan example, there is real uncertainty about the repayment: $Y$ is a random variable whose value cannot be predicted deterministically, and even the best prediction model cannot rule out this uncertainty. Similarly, in the COMPAS case, the fact of re-offending cannot be seen as a deterministic property of a delinquent. In all such cases, point estimators $\hat Y$ do not deliver useful information, and the only way of dealing with the uncertainty of the underlying situation is to use probabilities. This is reflected by Eq. (\ref{expected_utility}). 

Thus, from a decision-theoretical perspective, the basic task of the prediction-modeler is not to deliver a point estimate but a probability (even if there are cases where a point estimator may be useful), i.e., the required prediction model is a probabilistic prediction model. The ML task then consists of deriving an estimate $\hat p$ of the true probability, based on the analysis of historical data $\{\vec x_i, i=1,...n\}$, by specifying a function $f$ with $\hat p=f(\vec x)$. Since $f$ is determined from training data, it is prone to errors, and the resulting $\hat p$ is not identical to the real $p$. The goal of the prediction-modeler is thus to create a probability estimate that is as close as possible to the real $p$, as any deviation will lead to non-optimum decisions if the decision-maker uses the estimator $\hat p$ instead of the (unknown) true probability. 

If the decision rule is assumed to be given, this requirement can be somewhat relaxed: strictly speaking, the requirement is that  $\hat p$ leads to the same decisions as the true probabilities $p$. For example, in the loan context recalled in the last subsection, errors in $\hat p$ far away from the threshold specified in Eq. (\ref{optimum_solution_loan}) would not make any difference. Thus, in general, our framework is agnostic to the type of prediction model used. However, in all cases where the decision-making is not fully specified from the beginning, or the prediction-modeler does not have full access to all decision-relevant parameters, or the value of the decision-relevant parameters might change over time, the prediction-modeler has to care for generating a prediction model that works over the full range of $p$.

\subsection{The Interaction}
\label{ssec:The-interaction}

\begin{figure}[t]
    \centering
    \includegraphics[scale=0.5]{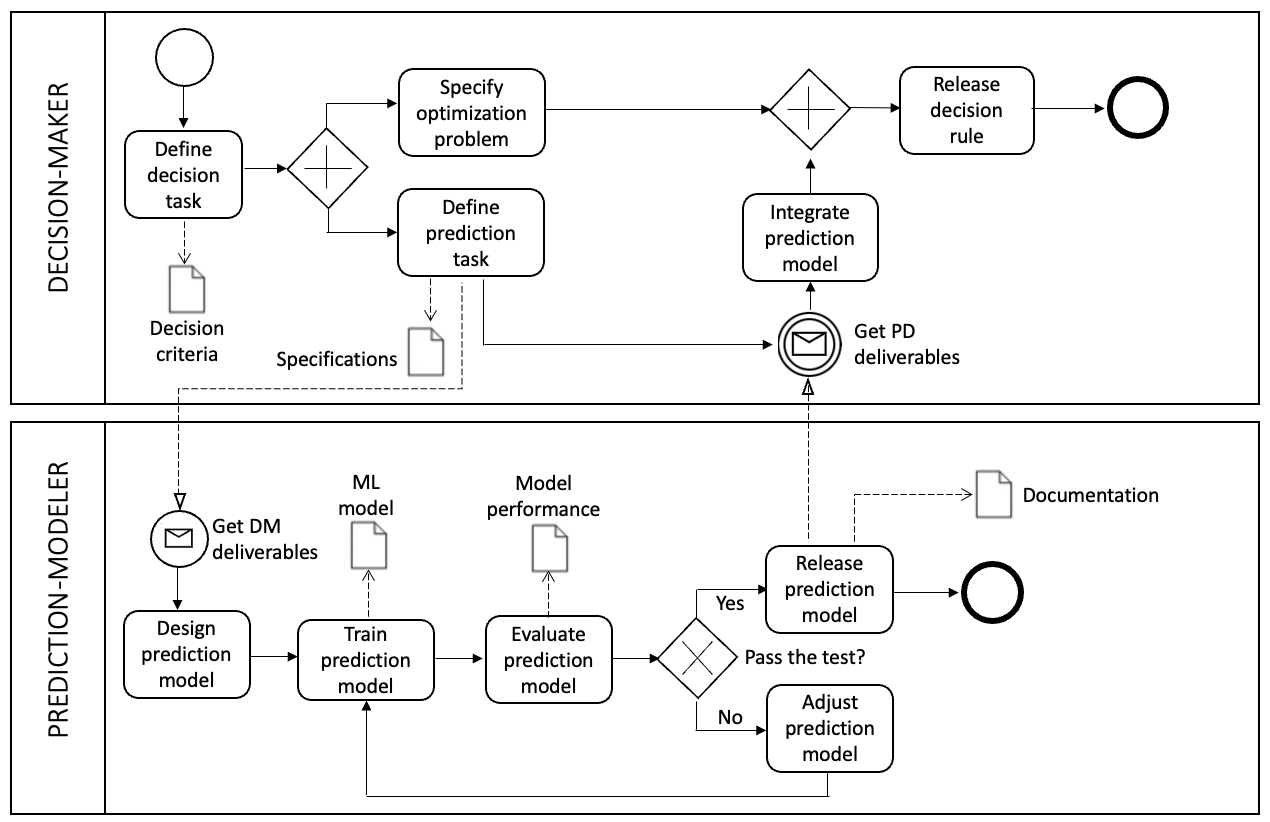}
    \caption{A BPMN (Business Process Model and Notation) diagram of the tasks involved in generating a prediction-based decision system.}
    \label{fig:BPMN_b}
\end{figure}

In this subsection, we analyze in more detail the interaction between the prediction-modeler and the decision-maker during the creation of a prediction-based decision system.
Figure~\ref{fig:BPMN_b} does this from a business process perspective.
In addition to the specific activities performed, we visualize the flow of information between the two roles required for developing a prediction-based decision system, focusing on the minimum interaction required between the two actors. The goal is to derive a minimum list of deliverables for this interaction and to study how these deliverables change when a fairness constraint is introduced. 

Table~\ref{table:minimum-deliverables} lists the minimum deliverables, i.e., the information that the decision-maker and the prediction-modeler must provide to each other, while Figure~\ref{fig:BPMN_b} visualizes \textit{when} during the sequence of tasks these deliverables are due. The decision-maker has to specify the prediction task according to the decision problem at stake. The prediction-modeler has to deliver a prediction model with associated additional information, such that the decision-maker can integrate it into the decision procedure. Depending on whether or not the decision-maker considers fairness requirements, the required deliverables differ, leading to two different scenarios for both actors, as depicted in Table~\ref{table:minimum-deliverables}: The row described by \textit{unconstrained utility maximization} refers to a decision-maker that bases their decision strictly on maximizing their utility without considering fairness (as formalized in Equation (\ref{optimize_utility})).
In contrast, the row \textit{utility maximization s.t. FC} refers to a decision-maker who also considers fairness, i.e., who adds a fairness constraint to the optimization problem, as is described in Equation (\ref{optimize_utility_fair}).
In the following, we will comment on and justify the elements of this table. 

\begin{table}
  \caption{Sets of minimum deliverables by role. \textbf{DM $\rightarrow$ PM} stands for the minimum set of deliverables the decision-maker (DM) must provide to the prediction-modeler (PM) and \textbf{PM $\rightarrow$ DM} describes the minimum set of deliverables the PM must provide to the DM.}
  \label{table:minimum-deliverables}
  \begin{tabular}
  {m{0.2\textwidth}m{0.27\textwidth}m{0.45\textwidth}}
    \toprule
        &\textbf{DM $\rightarrow$ PM}&\textbf{PM $\rightarrow$ DM}\\
    \midrule
        \multirow{3}{2.8cm}{unconstrained utility maximization} & $\bullet$ target variable $Y$     & $\bullet$ prediction model $\hat p=f(\vec x)$\\  
                                                            &                                   & $\bullet$ prediction model performance \\ 
                                                            &                                   & $\bullet$ calibration function \\ \hline
        \multirow{4}{2.8cm}{utility maximization s.t. FC}       & $\bullet$ target variable $Y$     & $\bullet$ prediction model $\hat p=f(\vec x)$\\  
                                                            & $\bullet$ sensitive attribute $A$ & $\bullet$ prediction model performance \\ 
                                                            &                                   & $\bullet$ group-specific calibration functions \\ 
                                                            &                                   & $\bullet$ group-specific baseline distributions\\ 
    \bottomrule
\end{tabular}
\end{table}

\subsubsection{Unconstrained Utility Maximization: DM $\rightarrow$ PM}
\label{sssec:DM_PM_unconstrained}
Defining the prediction task requires, at a minimum, the definition of the unknown variable $Y$ which should be predicted. In practice, additional specification elements such as the considered population or input features are given, which we omit for the sake of simplicity.%
\footnote{In certain situations, the prediction-modeler might receive training data from the decision-maker (for example, if it makes sense to use the decision-maker's customer data for training).
However, in other cases, the prediction-modeler relies on external data sources to develop a model predicting the target variable $Y$, as specified by the decision-maker.
Therefore, our general framework does not foresee the provision of training data by the decision-maker as a minimum deliverable.
}

\subsubsection{Unconstrained Utility Maximization: PM $\rightarrow$ DM }
\label{sssec:PM_DM_unconstrained}

Among the information provided by the prediction-modeler to the decision-maker, the prediction model is arguably the most important deliverable.
In addition to that, the prediction-modeler also needs to communicate the performance of the model. This is necessary for the decision-maker to assess whether the model fits the domain-specific requirements, i.e., to evaluate if the model should be included in the decision procedure or not. 

The decision-maker needs the probabilities $p$ to be able to derive the optimal decision rule (see  Equation (\ref{optimum_solution_loan})). However, many ML models deliver uncalibrated scores instead of an estimate of the probability. To fulfill the needs of the decision-maker, the prediction model needs to be calibrated, delivering an estimate $\hat p$ of $p$. If this is not the case, the prediction model should come with a calibration function that allows the decision-maker to reconstruct the probabilities from the score. Note that calibration defines ``a property of the model [more] than of its use since it does not depend on decision thresholds''~\cite[p.~55]{Hutchinson2019}.

\subsubsection{Utility Maximization s.t. FC: DM $\rightarrow$ PM }
\label{sssec:DM_PM_with_FC}

Consider now the minimum deliverables of the decision-maker in the constrained case, that is, when the decision-maker optimizes their utility subject to some group fairness constraint.
Recall that the basic idea of group fairness is to avoid unjustified disadvantages for subgroups of the population, defined by a sensitive attribute $A$ (see Section~\ref{ssec:Measuring-Group-Fairness}). The specification of the regarded sensitive attributes is done by the decision-maker. Only with knowledge of the protected subgroups considered, the prediction-modeler, in turn, can transmit the minimum deliverables assigned to them.

\subsubsection{Utility Maximization s.t. FC: PM $\rightarrow$ DM }
\label{sssec:PM_DM_with_FC}
In order to enable the decision-maker to solve the constraint optimization problem stated in Equation (\ref{optimize_utility_fair}), the prediction-modeler must deliver additional information. The type of information might depend on the fairness constraint, and while the problem has been studied for some cases of fairness constraints, the ML literature still has many unexplored areas. In the following, we restrict the discussion to the group fairness metrics that have been studied so far, relating to~\cite{hardt2016equality}, \cite{10.1145/3097983.3098095}, and~\cite{baumann2022sufficiency}.%
\footnote{
\cite{hardt2016equality} use a concept they call \textit{immediate utility} whereas~\cite{10.1145/3097983.3098095} use the concept of loss minimization for their proofs.
Both of these concepts can easily be translated to what we call the decision-maker's utility.
Therefore, their solutions hold for the constrained optimization problem, as we defined it.
The problem formalization of~\cite{baumann2022sufficiency} is in line with the constrained optimization problem, as we defined it.
}
\cite{hardt2016equality,10.1145/3097983.3098095} prove that any optimal decision rule $d^*$ that satisfies statistical parity, conditional statistical parity, equality of opportunity, or predictive equality takes the following form of group-specific thresholds, i.e.:
\begin{equation}
d^*=\begin{cases}
    1 & \text{$p \geq \tau_a$} \\
    0 & \text{otherwise}
\end{cases}
\label{optimal_decision_rules_under_fairness}
\end{equation}
where $\tau_a \in [0,1]$ denote different group-specific constants.%
\footnote{
Note that when choosing conditional statistical parity as the FC, these constants additionally depend on the ``legitimate'' attributes.
Furthermore, for the fairness criteria that combine two parity constraints (equalized odds and sufficiency), some randomization is needed~\cite{hardt2016equality,baumann2022sufficiency}.
For simplicity, we omit this for the rest of the discussion.
}
\cite{baumann2022sufficiency} prove that any optimal decision rule $d^*$ that satisfies predictive parity or false omission rate (FOR) parity takes the following form of group-specific upper- or lower-bound thresholds, i.e.:
\begin{equation}
\label{eq:optimal_decision_rule_group-specific_upper-or-lower-bound_thresholds}
d^{\ast}=\begin{cases}
    \begin{rcases}
    1, & \text{for $p \geq \tau_a$} \\
    0, & \text{otherwise}
    \end{rcases}
    \text{for $v > P(Y=1|A=a)$}\\
    \begin{rcases}
    1, & \text{for $p \leq \tau_a$} \\
    0, & \text{otherwise}
    \end{rcases}
    \text{for $v< P(Y=1|A=a)$}
\end{cases}
\end{equation}
where $v$ denotes the positive predictive value for the predictive parity fairness constraint -- the false omission rate in the case of FOR parity, respectively. $P(Y=1|A=a)$ denotes the prevalence of group $a$ (also called base rate), which is defined as the share of individuals belonging to the positive class.

The fairness constraint transforms into a condition relating the thresholds $\tau_0$ and $\tau_1$, where the exact form of this relation depends on the chosen fairness constraint. The decision-maker's utility is maximized by selecting the optimum one from all pairs $(\tau_0,\tau_1)$ defined by this relation, based on the resulting utility. To evaluate the utility, the distributions of $p$ for both groups are needed. This so-called ``baseline distribution'' describes how each subgroup is distributed over the probability range $p \in [0,1]$ (for details of the determination of optimum thresholds see~\cite{hardt2016equality,10.1145/3097983.3098095,baumann2022sufficiency}). For a given prediction model, the baseline distributions can be determined, at least approximately, from the training data, and this information has to be delivered to the decision-maker as a necessary element for their decision-making. Also, the utility evaluation can only be done if the calibration requirements are met on the level of the subgroups, so either the prediction model must be calibrated separately for each considered subgroup or group-specific calibration functions need to be provided.
We refer the interested reader to~\cite{BiasOnDemand2023FAccT}, who provide a set of simulated experiments demonstrating the effect of group-specific (baseline) distribution differences (i.e., various types of biases) on the prediction model, its performance, the resulting group-specific calibration functions, and the downstream fairness properties for given decision-making rules.

Thus, the fact that the decision-maker is considering fairness constraints leads to additional information requirements from the side of the prediction-modeler. Recall that our discussion is restricted to a selected set of group fairness criteria that have been previously studied. The deliverables corresponding to these criteria are specified in Table~\ref{table:minimum-deliverables}. For other fairness constraints, the additional requirements may be different. However, as a general rule, we might expect that imposing fairness constraints for the decision system generates additional information requirements that the prediction-modeler must meet. Simply delivering a black-box prediction model without this additional information is, in general, not sufficient for enabling the decision-maker to ensure a fair decision system. In Section~\ref{s:EthicalImplicationofOurFramework}, we will analyze the ethical consequences of this.

Note that the discussed examples in this subsection relate to the so-called post-processing methods for creating fairness~\cite{mehrabi2019survey}, assuming that a decision-maker accepts the prediction model as given. This is the simplest situation with minimum interaction between the two players. However, our framework (as presented in Fig.~\ref{fig:BPMN_b}) also holds in cases where pre-processing or in-processing methods are applied. In such cases, the interaction between the two roles is more complicated, as the decision-maker has to inform the prediction-modeler about the fairness constraint and, at least for in-processing methods, specify the decision rule upfront. Thus, the task of generating fairness can be shifted to the prediction-modeler, but at the expense that the decision-maker restricts their freedom to change the decision rule after the prediction model is delivered. Thus, pre-processing and in-processing approaches require a closer collaboration of the two roles, with associated increased requirements for the interaction between the two roles.

\section{Discussion}
\label{s:EthicalImplicationofOurFramework}

The first important insight is that different actors come with different responsibilities. Here, we focus more specifically on professional responsibility, that is, the set of obligations based on a role played in a certain context.\footnote{Responsibility, of course, extends beyond roles, and for a broader discussion see~\cite{dePoel2011}.} Since our analysis relates to a well-defined problem in algorithmic decision-making (i.e., group fairness), these obligations translate into specific pieces of information that each role is expected to deliver. 

The deliverables we suggested are not optional and reflect the strong interdependence between roles. Ultimately, we acknowledge that the responsibility for fair decisions falls on the role of the decision-maker for the reasons we already discussed in Section~\ref{s:Prediction-Based-Decision-Framework}. However, their ability to address fairness issues depends heavily on the work of the prediction-modeler. Similarly, the latter cannot take responsibility for group-specific calibration functions and baseline distributions if they do not receive information about the sensitive attributes to be considered.

The interdependence between roles recalls the problem of creating meaningful communication channels among designers, managers, and, more generally, all professionals involved in designing and using artificial intelligence (AI) systems. To this aim, we offer some considerations that might be useful to inform future research and the implementation of prediction-based decision-making systems.

So far, most of the literature on algorithmic fairness underestimated practical issues emerging in real-world organizations (see~\cite{holstein2019improving} for a notable counterexample), but to provide effective and sustainable solutions, we need to fill the gap between mathematical abstractions and organizational dynamics and engineering practices~\cite{aler2022ethical}. Scholars have already addressed the risks of abstracting from the social context of AI applications and highlighted the need to reorient technical efforts from solution-oriented approaches to process-oriented ones~\cite{selbst2019fairness, scantamburlo2021non}. Our framework goes in that direction and tries to figure out which kind of concrete interactions would help the implementation of (group) fairness, starting from two key roles and their associated tasks.
In the future, these processes will have to be supported by appropriate tools, such as Fairlearn~\cite{bird2020fairlearn}, AI360~\cite{AIF365}, or the FairnessLab~\cite{hertweck2023fairnesslab}, which have recently been developed to help develop more fair prediction-based decision-making systems.

Starting from professional roles gives the opportunity to distill important information entering prediction and decision tasks and directs greater attention to organizational aspects, which are often less regarded in the field of AI ethics. In general, analyzing roles and their interactions can reveal the background of values and assumptions that shape the design process~\cite{krijger2021enter}. 
This role-based perspective may also serve to highlight a more articulated view of the design and use of algorithmic decision-making systems, where more than one professional might be involved. Usually, the study of human-AI interaction focuses on the exchange occurring between the (end) users and the operating systems. However, our framework suggests that there are other meaningful interactions that are worthy of consideration. An analysis of interactions shaping the design and use of AI systems may reveal conceptual gaps, structural deficiencies, and power imbalances.  

Our exercise considers a simple business process scenario, but other elaborations are possible (a finer-grained analysis of tasks in different settings, e.g., medicine). For example, further research might explore connections with existing frameworks that emphasize the context-sensitive nature of computing systems, such as the model of contextual integrity~\cite{nissenbaum2010}. A closer look at the norms and social practices that control, manage, and steer the flow of information within organizations can help gain a richer understanding of prediction-based decision systems. This may result in a description of the flows of information characterizing the context of a prediction-based decision system and the identification of which flows are appropriate to ensure agents (and the organizations) meet established goals and ethical norms.

For supporting interactions among professionals, an essential task is to keep track of relevant information characterizing the elements of the algorithmic decision-making system. In computer science and engineering disciplines, this goal is often fulfilled by devising software documentation that may include a variety of information (e.g., technical requirements, software architecture, codes, etc). Note that documentation is also acknowledged as an important measure to ensure transparency and accountability of AI systems. In this regard, the European Commission's proposal for an AI regulation requires that ``technical documentation of a high-risk AI system shall be drawn up before that system is placed on the market or put into service and shall be kept up-to-date'' (article 11~\cite{AiAct2021}). However, it is still unclear which types of information should form a robust documentation. This effort, moreover, should also consider how to make information accessible and useful for the players contributing to the informational exchange. This would require addressing issues of knowledge and language divides, which often characterize participatory design processes. 

With respect to the documentation task, the results suggested by our framework have a limited scope in that they refer to a particular setting (i.e., group fairness in algorithmic decision-making). However, our results show that, in general, it is necessary to perform an analysis of which specific contents one may need to address the problem at stake. Our effort suggests that, in general, players might have to ask for more specific information due to the context of use and the ethical issues addressed. Also, our attempt shows that it is necessary to address the question of how to modulate the creation and maintenance of documentation among different players. So far, technical documentation is often conceived as a task entirely in charge of engineers and computer scientists, but, in reality, there might be other roles affecting the design and the deployment of AI and ML systems. So we might think of reporting and documenting more as a collaborative practice that involves different roles rather than a duty assigned to a single category of people.

A final consideration regards human oversight, an ethical principle recommending human agency in AI-driven decision processes to ensure human autonomy and prevent adverse effects. 
While the notion of human-in-the-loop can inspire the structuring of human intervention and monitoring, it is open to discussion of what type of duties and actions would be needed in real-world scenarios: What does it mean to intervene in a decision cycle? Who should do it? The intuition of identifying roles and the associated tasks is a way to start answering such questions. This would be particularly beneficial because in real-world decision-making procedures (such as those embedded in administrations or bureaucratic processes) responsibility is often delegated and distributed across multiple actors~\cite{strandburg2021}. In our framework, we envision activities and interactions based on a simplified Business Model Notation, but richer elaboration can provide more details on who supervises what. 

The creation of a flow of information between the prediction-modeler and the decision-maker connects to key ethical requirements in the design and deployment of AI and ML systems: Transparency, accountability, and human oversight. The scientific community and policymakers largely acknowledge the centrality of these requirements. However, there is still limited knowledge and experience on translating these requirements into practice. The approach our framework suggests offers meaningful stimuli to articulate these requirements more concretely and raises points that can move the community towards new research and policy directions.

\section{Conclusions}

In this paper, we argue that a prediction model as such cannot be qualified as fair or unfair. This argument is based on two observations: First, predictions themselves have no direct impact. Second, predictions can be used differently for making decisions. Important examples for the second observation are all post-processing methods to implement fairness constraints, e.g.,~\cite{baumann2022sufficiency,hardt2016equality,10.1145/3097983.3098095}. These methods are based on the idea that the fairness properties of a decision system can be shaped by the way in which the prediction model's output is transformed into a decision, e.g., by imposing group-dependent decision thresholds. So, the same prediction model can lead to unfairness (without post-processing) or fairness (with adequate post-processing). Other examples are all human-in-the-loop approaches that combine prediction models with human decision-makers. They assume that humans are at least co-creators of the resulting ethical consequences of prediction-based decision systems, which of course implies that different ways of using the prediction model's output are conceivable and that the activity of the human in the loop consists exactly in influencing the usage of the prediction model's output.

Our framework represents an exercise that allows us to pinpoint what is essential for each role when addressing fairness in a prediction-based decision system. This lets us suggest a minimum level of active responsibility~\cite{dePoel2011} that one could demand from these roles in similar situations. 
The identification of deliverables and interactions is not meant to limit the responsibility of ML developers and decision-makers to the delivery of specific pieces of information, but to avoid false or too vague expectations of the obligations for the roles involved. 
Actually, a better understanding of the different roles involved with their associated goals, tasks, and professional responsibilities, is an important first step to take care of the implementation of ethical requirements in prediction-based decision-making systems. 

\begin{acks}
TS was supported by by the project ``A European AI On Demand Platform and Ecosystem'' (AI4EU) -- H2020-ICT-26 \#825619.
JB and CH were supported by the National Research Programme ``Digital Transformation'' (NRP 77) of the Swiss National Science Foundation (SNSF) -- grant number 187473.
\end{acks}

\bibliographystyle{plain}
\bibliography{references}

\end{document}